# Surface Charge Induced Dirac Band Splitting in a Charge Density Wave Material (TaSe$_4$)$_2$I

Hemian Yi[1], Zengle Huang[2], Wujun Shi[3], Lujin Min[1], Rui Wu[4], C. M. Polley[5], Ruoxi Zhang[1], Yi-Fan Zhao[1], Ling-Jie Zhou[1], J. Adell[5], Xin Gui[6], Weiwei Xie[7], Moses H. W. Chan[1], Zhiqiang Mao[1], Zhijun Wang[8,9], Weida Wu[2], and Cui-Zu Chang[1]

[1]Department of Physics, The Pennsylvania State University, University Park, PA 16802, USA

[2]Department of Physics and Astronomy, Rutgers University, Piscataway, NJ 08854, USA

[3]School of Physical Science and Technology, ShanghaiTech University, Shanghai 200031, China

[4]Beijing Academy of Quantum Information Sciences, Beijing 100193, China

[5]MAX IV Laboratory, Lund University, 221 00 Lund, Sweden

[6]Department of Chemistry, Princeton University, Princeton, NJ, 08540, USA

[7]Department of Chemistry and Chemical Biology, Rutgers University, Piscataway, NJ 08854, USA

[8]Beijing National Laboratory for Condensed Matter Physics and Institute of Physics, Chinese Academy of Sciences, Beijing 100190, China

[9]University of Chinese Academy of Sciences, Beijing 100049, China

Corresponding author: cxc955@psu.edu (C.-Z. C.)

**Abstract:** **(TaSe$_4$)$_2$I, a quasi-one-dimensional (1D) crystal, shows a characteristic temperature-driven metal-insulator phase transition. Above the charge density wave (CDW) temperature $T_c$, (TaSe$_4$)$_2$I has been predicted to harbor a Weyl semimetal phase. Below $T_c$, it becomes an axion insulator. Here, we performed angle-resolved photoemission spectroscopy (ARPES) measurements on the (110) surface of (TaSe$_4$)$_2$I and observed two sets** **of Dirac-like energy bands in the first Brillion zone, which agree well with our first-principles calculations. Moreover, we found that each Dirac band exhibits an energy splitting of hundreds of meV under certain circumstances. In combination with core level measurements, our theoretical analysis showed that this Dirac band splitting is a result of surface charge polarization due to the loss of surface iodine atoms. Our findings here shed new light on the interplay between band topology and CDW order in Peierls compounds and will motivate more studies on topological properties of strongly correlated quasi-1D materials.**

**Main text:** Topological semimetals, which are characterized by linear band crossings near the Fermi level, have attracted intense interest in the condensed matter physics community in the last decade. Based on the physical origin of the band crossing, topological semimetals are categorized into Dirac semimetals, Weyl semimetals, nodal-line semimetals, and others [1,2]. Due to the spin degeneracy, the Dirac node in Dirac semimetals is composed of two massless Weyl nodes with opposite chirality, which overlap with each other at the same momentum. By breaking either time-reversal or inversion symmetry, the Dirac band can split into a pair of Weyl bands. The Dirac semimetal state was first discovered in $Na_3Bi$ [3] and $Cd_3As_2$ [4-6]. The non-centrosymmetric Weyl semimetal state was first observed in TaAs [7-11], while the magnetic Weyl semimetal state was recently discovered in $Co_3Sn_2S_2$ [12,13] and $Co_2MnGa$ [14]. The observation of the Dirac/Weyl nodes and/or the surface Fermi arcs in angle-resolved photoemission spectroscopy (ARPES) measurements have been considered as direct evidence for these topological semimetals. Prior studies on topological semimetals have focused on three dimensional (3D) crystals, which can be theoretically described by the single-particle model. More recently, the Weyl semimetal phase coupled with the charge density wave (CDW) order has been predicted [15,16] and recently claimed to be observed [17] in a correlated quasi-one-dimensional (1D) halogened transition metal tetrachalcogenide (TaSe$_4$)$_2$I crystal.

(TaSe$_4$)$_2$I is a prototype quasi-1D Peierls compound [18]. Its conventional unit cell consists of two TaSe$_4$ chains parallel to the *c*-axis, which are separated by the iodine (I) atom planes (Fig.1a). In the TaSe$_4$ chain, each Ta atom is sandwiched by two rectangular selenium units. The dihedral angle between adjacent rectangles is 45°, so this unique stacking order makes TaSe$_4$ chains show a screw-like symmetry and thus (TaSe$_4$)$_2$I is known as a chiral crystal [19]. (TaSe$_4$)$_2$I undergoes a Peierls transition at $T \sim 263$ K [17,20]. Above the CDW temperature $T_c = 263$ K, first-principles calculations show (TaSe$_4$)$_2$I to be a Weyl semimetal, originating from the chiral structure induced inversion symmetry breaking. Below $T_c$, the CDW gap of hundreds of meV can drive (TaSe$_4$)$_2$I into an axion insulator state [15-17]. Unlike the axion insulator state realized in molecular beam epitaxy grown magnetic topological insulator sandwich heterostructures, where the gap opening at Dirac point is induced by internal magnetization [21-23], the axion insulator realized here arises from the formation of the CDW order in a Weyl semimetal [24-26]. One recent transport study on (TaSe$_4$)$_2$I claimed that the axion insulator state appears at low temperatures [17]. Prior ARPES studies on (TaSe$_4$)$_2$I have interpreted the gap near the Fermi level as a polaron gap, while the CDW gap opens away from the Fermi level [27,28]. The "V" shaped band structure observed along the $\overline{\Gamma Z}$ direction (Fig.1b) has been claimed to be induced by competing periodic potentials [29]. To date, the linearly dispersed energy bands, and their electronic nodes have not been resolved in (TaSe$_4$)$_2$I.

In this *Letter*, we reported the observation of two sets of Dirac-like energy bands along the $\overline{\Gamma Z}$ direction, revealed through synchrotron ARPES measurements on the (TaSe$_4$)$_2$I (110) surface. This result agrees well with our first-principles calculations. We further found that the Dirac band exhibits an energy splitting of hundreds of meV under certain circumstances. The circular dichroism (CD) effect of the Dirac band splitting was also studied, which validates the observed Dirac band splitting to be an intrinsic property of $(TaSe_4)_2I$. We showed that the band splitting observed here is not the spin splitting of Dirac bands into Weyl bands [15], which is beyond the resolution of our ARPES. By combining the core level measurements and theoretical calculations, we demonstrated that the Dirac band splitting is a result of surface charge polarization due to the loss of iodine atoms. The observation of the electronic nodes and the Dirac band splitting will help us comprehensively understand the correlated topological properties in this CDW semimetal $(TaSe_4)_2I$.

The $(TaSe_4)_2I$ single crystals used in this work were synthesized using the chemical transport method [30], with a typical size of $0.8 \times 0.8 \times 5$ mm$^3$. The $(TaSe_4)_2I$ single crystals were characterized by the (*n n* 0) scattering peaks from X-ray diffraction (XRD) measurements (Fig. S2c)[31]. Our single-crystal XRD results validate the chiral structure of $(TaSe_4)_2I$ (Fig. S1 and Tables S1 to S3)[31]. The fresh (110) surface is unveiled by cleaving the $(TaSe_4)_2I$ crystal in the ultrahigh vacuum chamber with a base vacuum better than $5 \times 10^{-11}$ mbar. The ARPES measurements were performed at the Bloch beamline of MAX IV (Sweden) and beamline 5-2 of SLAC (USA). The CD-ARPES measurements were performed at the SIS beamline of SLS (Switzerland). The hemispherical Scienta DA30L analyzer is used in all our ARPES measurements. The energy and angle resolution were set to ~10 meV and ~0.1°, respectively. The scanning tunneling microscopy (STM) experiments were carried out in an Omicron low temperature (LT)-STM system. Our first-principles calculations were performed based on density functional theory using the full-potential local-orbital basis within the generalized gradient approximation. The spin-orbit coupling (SOC) effect is included in all our calculations.

We first used the LT-STM to characterize the collective CDW phase in (TaSe$_4$)$_2$I. Figures 1c and 1d show the atomically resolved STM images acquired on the freshly cleaved (TaSe$_4$)$_2$I (110) surface at about 150 K (below $T_c$ ~ 263 K). The bias dependence of STM images indicates the cleaved surface is the iodine (I) atom layer, in which ~50% iodine atoms remain (Fig. 1d). A few iodine vacancies are randomly distributed on the (110) surface. The Fourier transform of the STM image (Fig. 1c inset) shows two sharp superlattice spots of the CDW modulations. Long-range CDW stripes with periodicity ~17 ± 1 nm at 150 K are observed on the entire scanned area (Fig. 1c), which is longer than that (~10.6 nm) from prior X-ray diffraction and neutron scattering measurements [32-37]. This difference might come from the change of nesting condition because of different carrier concentrations. The orientation of the CDW modulation is ~ 45° ± 0.5° off the $c$-axis of the crystal (along the TaSe$_4$ chains), indicating the CDW wavevector $\vec{q}_{CDW}$ = (-0.027, 0.027, 0.054) is in the (110) plane. The $\vec{q}_{CDW}$ value is similar to that in prior studies [32-37]. The stripe modulation is absent at room temperature (above $T_c$), further confirming the occurrence of the CDW transition in (TaSe$_4$)$_2$I [31]. The CDW phase transition of (TaSe$_4$)$_2$I is also revealed in our four-terminal electrical transport measurements. Figure 1e shows the logarithmic normalized longitudinal resistivity $\rho/\rho(T = 300\ K)$ and its derivative as a function of $10^3\ T^{-1}$. A pronounced peak is observed at the CDW transition temperature $T_c$ ~263 K, consistent with prior transport results on this Peierls material [17,20].

Next, we performed ARPES measurements on the $\overline{Y\Gamma Z}$ and $\overline{X\Gamma Z}$ planes of the freshly-cleaved (TaSe$_4$)$_2$I (110) surface(Fig. 1b). We observed a periodic feature along the $k_z$ direction in the constant energy contour (CEC) mapping, which is acquired on the $\overline{Y\Gamma Z}$ plane at $E$-$E_F$ = -0.6 eV (Fig. S5b)[31]. The appearance of the oppositely bent stripe band structures confirms its 3D property, i.e., they originate from the bulk valence bands rather than the surface states of (TaSe$_4$)$_2$I.

The CEC mapping acquired on the $\overline{X\Gamma Z}$ plane at $E$-$E_F$ = -0.6 eV and the electronic band spectra along the $\overline{\Gamma Z}$ direction measured using the 23 eV light with linear horizontal (LH) polarization are shown in Figs. S5c and S5d. The "M"-shaped band structure implies the lower branch of the Dirac band. The anisotropic electronic structures in (TaSe$_4$)$_2$I is consistent with its quasi-1D crystal structure with two interacted neighboring TaSe$_4$ chains in one conventional unit cell [31].

To further demonstrate the complete Dirac band crossings, we performed ARPES measurements along the $\overline{\Gamma Z}$ direction using the 23 eV light with linear vertical (LV) polarization at room temperature. Two sets of Dirac-like energy bands are resolved, which agree well with our calculated electronic band structures (Fig. 2a). The Dirac nodes are located at $E$-$E_F$ = -0.39 eV, consistent with the electron carriers derived from prior Hall measurements [17]. The two Dirac nodes are seen clearly in the momentum distribution curves (MDC) of the normalized ARPES band spectra (Fig. 2b). A pair of separated Weyl nodes have been predicted to appear along $\overline{\Gamma Z}$ direction above the CDW temperature $T_c$ [15]. Since this splitting is on the order of few meV, neither these two separated Weyl nodes nor the predicted Fermi surface segments on (TaSe$_4$)$_2$I (110) surface are resolved in our ARPES experiments. It is worth noting that the two Dirac nodes are located at $k_F \sim \pm 0.27/$ Å$^{-1}$, which is equal to ~1.1 $\pi/c$ ($c$ ~ 12.778 Å). This value is slightly larger than half of the CDW wavevector (~ 1.055 $\pi/c$) calculated by our STM results [31], consistent with the heavily electron-doped property of (TaSe$_4$)$_2$I. $\vec{q}_{CDW}$ ~ 2.11 $\pi/c$ serves as the nesting vector at the Fermi surface in the CDW phase of (TaSe$_4$)$_2$I [28].

With the electronic nodes identified, we studied the temperature evolution of the electronic band structure along $\overline{\Gamma Z}$ direction. We heat the freshly cleaved (TaSe$_4$)$_2$I sample to $T$ = 350 K and later cool the samples down to $T$ = 265 K and 120 K, respectively. At $T$ =350 K, there are no coherent band spectra observed across the Fermi level (Fig. 3a). A gap of ~ 120 meV opens at the Fermi level at $T = 265$ K (Fig. 3b) and it increases (~ 240 meV) at $T = 120$ K (Fig. 3c). We speculate this gap is induced by the formation of the CDW state in the sample in addition to the presence of the polaron excitations as claimed in a prior report [27]. Based on the Peierls scenario, it is unlikely the CDW gap opens at the bulk conduction/valence band crossing points instead of the Fermi level [18]. The larger gap observed at a lower temperature can be seen from the onset of the energy distribution curve (EDC) of these ARPES spectra. A fixed peak is observed at the Γ point, which excludes the possibility that the gap opening is induced by the band shift. The bulk valence band shows the reversed "V" shape feature at all temperatures, i.e. the intersecting point is always located at the EDC onset and moves downward with decreasing temperature. In addition, the gap opened near the Fermi level appears when $T > T_c$. These two aspects indeed imply that the polaron gap might coexist with the CDW gap, both of which open near the Fermi level. More experimental and theoretical studies are needed to clarify this issue.

In addition to the gap opened near the Fermi level, we observed a clear Dirac band splitting in these ARPES spectra acquired at $T = 350$ K, 265 K, and 120 K (Figs. 3a to 3c). With decreasing temperature, the Dirac band splitting becomes smaller. We noted that no Dirac band splitting has been observed in the room temperature ARPES band maps shown in Fig. 2. Since the sample studied here has been heated to $T = 350$ K, we speculate the loss of iodine atoms on the $(TaSe_4)_2I$ (110) surface is responsible for the Dirac band splitting. Therefore, we performed core level measurements, which can characterize the loss of iodine atoms, and corresponding ARPES measurements on this sample. We found that this Dirac band splitting is indeed related to the iodine atom loss induced surface charge in $(TaSe_4)_2I$ samples (Figs. S6 and S7) [31]. The Dirac band splitting is much clearly resolved in the second derivative spectra of the ARPES bands (Fig. S9a to S9c) [31]. To quantitatively study the band splitting as a function a temperature, we extracted EDCs at the same momentum $k \sim 0.246$ Å$^{-1}$ (i.e. $\pi/c$). The two dip curves of each EDC are fitted by two Gaussian functions. The energy offsets between the two split bands are 0.31 eV, 0.22 eV, and 0.16 eV, respectively (Fig. S9d) [31]. The proposed Weyl band splitting on (TaSe$_4$)$_2$I (110) surface is on the order of few meV, which is much smaller than the linearly dispersed band splitting observed here [15]. Moreover, the Weyl band splitting is usually independent of temperature. Therefore, we concluded the Dirac band splitting observed in our experiment is not the expected Weyl band slitting on (TaSe$_4$)$_2$I (110) surface [15].

To exclude the possibility that the Dirac band splitting observed here is from band spectra of two different (TaSe$_4$)$_2$I pieces, we performed CD-ARPES measurements. The CD effect in ARPES has been widely used to examine the texture of spin and/or orbital angular momentums in the Dirac surface states of topological materials [38-42]. Figures 3d and 3e show the ARPES band mapping along $\overline{\Gamma Z}$ direction obtained using the right circularly polarized (RCP) and left circularly polarized (LCP) lights, respectively. The photoemission intensity distributions of these two spectra show a clear dependence of the circular polarization of the incident photon. For instance, we observed four electronic nodes in the second Brillion zone in Fig. 3e, some of which are absent in Fig. 3d. To better resolve the CD effect, we calculated the magnitude of normalized CD intensity distribution using the following equation: CD = ($I_{RCP} - I_{LCP}$)/ ($I_{RCP} + I_{LCP}$), where $I_{RCP}$ ($I_{LCP}$) is the intensity of photoemission measured using the RCP (LCP) light. The observation of the alternating red-blue intensity distribution further proves the existence of the CD effect in the electronic structure of (TaSe$_4$)$_2$I (Fig. 3f). The CD effect can also be seen from the CEC mapping acquired on the $\overline{X\Gamma Z}$ plane at $E$- $E_F$ = -1.06 eV (Figs. S10b and S10c) [31]. These results suggest (*i*) the CD intensity of two Dirac bands located at $\pm$ 0.27 Å$^{-1}$ shows the left-right antisymmetric, and (*ii*) the sign of the CD intensity distribution is also reversed for these two sets of Dirac bands. Therefore, the observed Dirac band splitting is an intrinsic property of (TaSe$_4$)$_2$I.

To understand the origin of the Dirac band splitting observed in (TaSe$_4$)$_2$I, we performed first-principles calculations on (TaSe$_4$)$_2$I on (110)-terminated slab structures with different amounts of iodine atoms on (110) surface. Since the cleaving process occurs at the iodine termination, the iodine atoms can easily leave from the cleaved surface after heating. Figures 4a to 4c show the calculated electronic band structures with 100%, 50%, 0% residual iodine atoms on (TaSe$_4$)$_2$I (110) surface, respectively. The surface spectra are obtained by projecting the energy bands of the slab calculation onto different surface terminations. In Figs. 4a to 4c, the size of the red (blue) dots represents the fraction of electronic charge on the outermost (second outermost) TaSe$_4$ chains, while the size of the grey dots represents the electronic charge on the rest bulk atoms.

For an ideal cleaving of the crystal plane, 50% of iodine atoms may stay on the (110) surface (i.e. charge neutrality) (Fig.1d), and the Dirac bands coincide with the bulk bands and no band splitting appears (Fig. 4b). For a cleaved (110) surface with 100% iodine atoms and without iodine atoms (i.e. surface charge polarization), significant upward and downward shifts of surface Dirac bands are seen (Figs. 4a and 4c). We, therefore, attributed the Dirac band splitting observed in our ARPES measurements (Fig. 3) to the formation of surface charge due to the loss of iodine atoms in the surface layers. While 50% iodine surface maintains electrical neutrality, more or fewer iodine atoms at the surface would introduce a static electronic field, which is perpendicular to the surface. The static electric field could separate the surface Dirac states from the bulk states. More (less) adsorbed iodine atoms would drive the outermost Dirac bands to higher (lower) energy. We noted that the opposite CD effect observed in the outermost and second outermost (TaSe$_4$)$_2$I layers (Figs. 4f and S10c) might be a result of the $\frac{c}{2}$ shift between these two layers and/or the appearance of the static electric field-induced Rashba-type band splitting in the $(TaSe_4)_2I$ surface layers. Finally, we can also explain why the magnitude of the Dirac band splitting becomes smaller with decreasing temperature (Fig. 3a to 3c): Our transport results show that the sample becomes more insulating at lower temperatures (Fig. 1e), this will weaken the screening effect and reduce the electrical potential gradient between the outermost and second outermost layers and thus lead to a smaller Dirac band splitting.

In summary, we observed the Dirac-like bands in the first Brillion zone of $(TaSe_4)_2I$ (110) surface, consistent with our first-principles calculations. By heat-treating freshly cleaved $(TaSe_4)_2I$ (110) surface, we found Dirac band splitting, which based on our core level measurements and theoretical analysis is attributed to the loss of iodine atoms on the surface. The observed Dirac band splitting show a strong circular dichroism effect, validating that the Dirac band splitting observed in $(TaSe_4)_2I$ is an intrinsic property. Our findings shed light on the interaction between the Dirac bands and CDW order in Peierls compounds and provide new insights in the field of strongly correlated topological materials.

**Acknowledgments**

The authors would like to thank C. X. Liu, Z. Wang, B. H. Yan, D. Xiao, Y. T. Cui, J. Yan, and X. D. Xu for helpful discussion. This work is primarily supported by the NSF-CAREER award (DMR-1847811), including the ARPES measurements, electrical transport measurements, and data analysis. The sample characterization is partially supported by the DOE grant (DE-SC0019064). C. Z. C. also acknowledges the support from the Gordon and Betty Moore Foundation's EPiQS Initiative (Grant GBMF9063 to C. Z. C.) and the Penn State NSF-MRSEC grant (DMR-2011839). L.H.M. and Z.Q.M. acknowledge the support by the Penn State NSF-MRSEC grant (DMR-2011839). W.W. and Z.H. acknowledge the support from ARO award (W911NF-20-1-0108).

**Figures and figure captions**

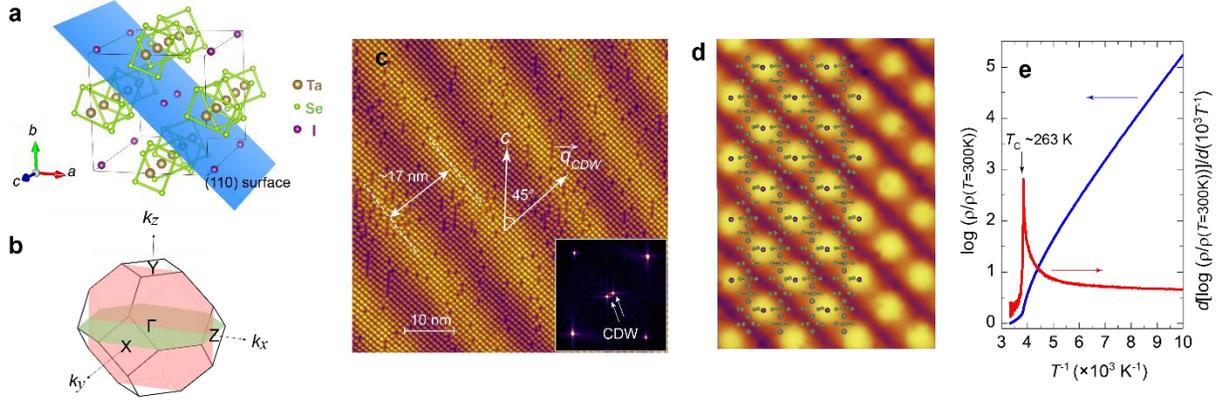

**Figure 1 | Crystal structure and sample characterization. a,** Crystal structure of (TaSe$_4$)$_2$I. The (110) surface is shown in blue. **b,** Bulk Brillouin zone of (TaSe$_4$)$_2$I. **c,** STM image of (TaSe$_4$)$_2$I (110) surface. The $c$ axis of the crystal and the CDW wavevector $\vec{q}_{CDW}$ are shown. This image is acquired at -1 V/30 pA. Inset: the Fourier transform of the STM image. **d,** Enlarged STM image of the green dashed rectangle area shown in (b). ~50% of iodine atoms are expected to remain on the freshly cleaved (110) surface. **e,** The logarithmic normalized longitudinal resistivity ρ/ρ($T$ =300 K) (blue) and its derivative (red) as a function of $10^3\ T^{-1}$.

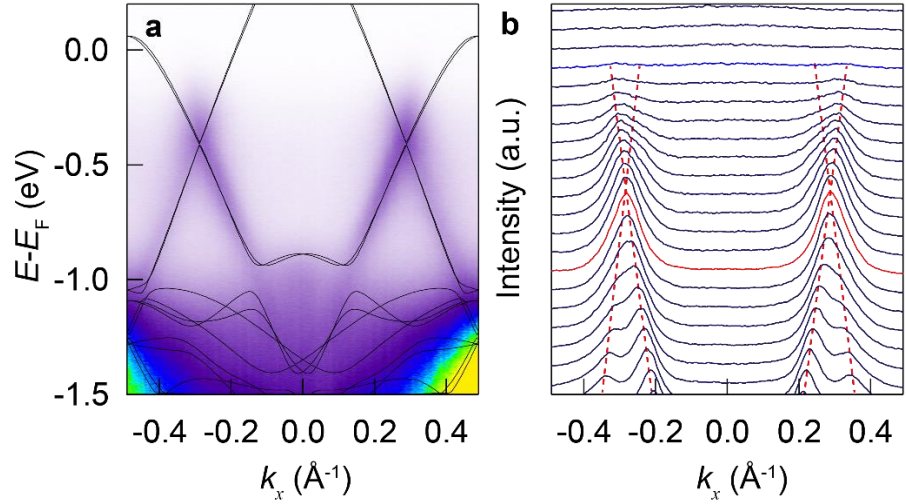

**Figure 2 | Observation of Dirac bands on (TaSe$_4$)$_2$I (110) surface. a,** ARPES spectra along the $\overline{\Gamma Z}$ direction. The 23eV light with LV polarization is used. The black curves are the calculated band structures. **b,** MDC of the normalized ARPES band spectra.

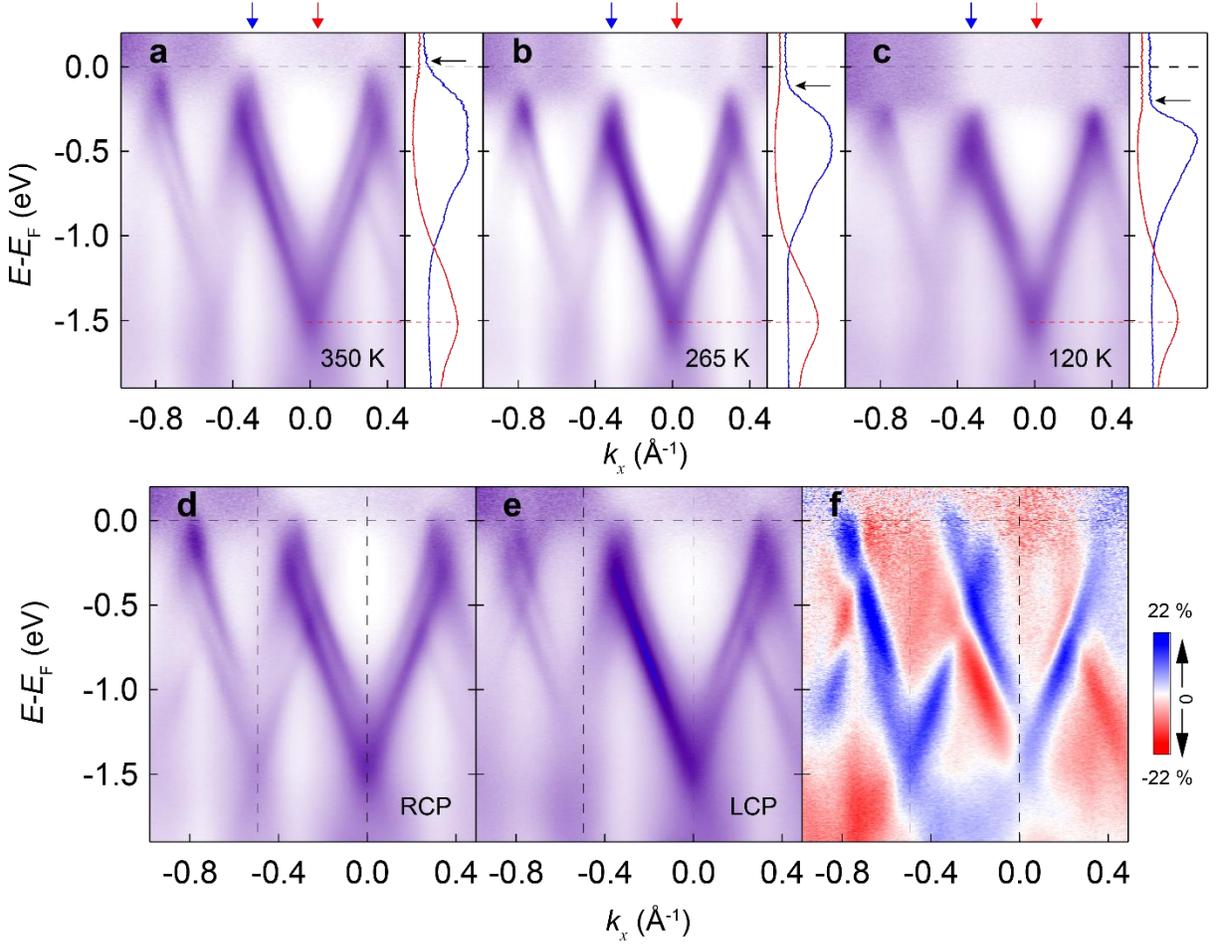

**Figure 3 | Observation of Dirac band splitting on heat-treated (TaSe$_4$)$_2$I (110) surface. a-c,** ARPES spectra along the $\overline{\Gamma Z}$ direction acquired at $T$ = 350 K (a), 265 K (b), and 120 K (c). Right panels of (a-c): two EDC lines extracted at two momenta marked with red and blue arrows above ARPES spectra. The 47eV light with LH polarization is used in (a-c). **d, e,** ARPES spectra along the $\overline{\Gamma Z}$ direction measured with LCP and RCP 47eV lights, respectively. **f,** Normalized CD spectra along the $\overline{\Gamma Z}$ direction. ARPES data shown in (**d** and **e**) are taken at $T$ = 350 K.

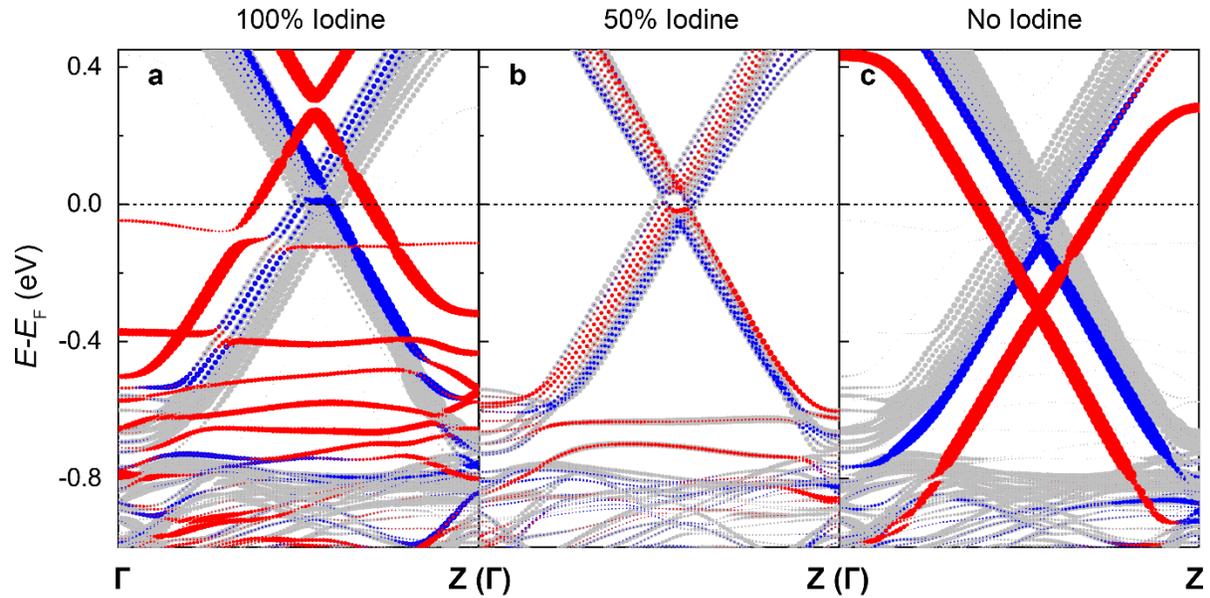

**Figure 4 | Surface charge induced Dirac band splitting on (TaSe$_4$)$_2$I (110) surface. a-c,** Calculated band structures of (TaSe$_4$)$_2$I along the ΓZ direction with the integrated $k_z$. 100% (a), 50% (b), and 0% (c) of iodine atoms on (TaSe$_4$)$_2$I (110) surface. The size of the red (blue) dots represents the fraction of electronic charge on the outermost (second outermost) TaSe$_4$ layers, while the size of the grey dots represents that of the rest bulk atoms.